\documentclass[iop]{emulateapj}
\usepackage{apjfonts}

\newcommand{\spitzer}{{\it Spitzer}}
\newcommand{\hst}{{\it HST}}
\newcommand{\corot}{{\it CoRoT}}
\newcommand{\kepler}{{\it Kepler}}
\newcommand{\um}{\,$\rm \mu m$}

\shorttitle{Thermal emission of WASP-1b \& WASP-2b}
\shortauthors{Wheatley et al.}

\begin{document}


\title{The thermal emission of the exoplanets 
WASP-1\lowercase{b} and WASP-2\lowercase{b}}


\author{
Peter~J.~Wheatley\altaffilmark{1}, 
Andrew~Collier~Cameron\altaffilmark{2},
Joseph~Harrington\altaffilmark{3}, 
Jonathan~J.~Fortney\altaffilmark{4},
James~M.~Simpson\altaffilmark{2},
David~R.~Anderson\altaffilmark{5},
Alexis~M.~S.~Smith\altaffilmark{5},
Suzanne~Aigrain\altaffilmark{6},
William~I.~Clarkson\altaffilmark{7},
Micha\"el~Gillon\altaffilmark{8,9},
Carole~A.~Haswell\altaffilmark{10},
Leslie~Hebb\altaffilmark{11},
Guillaume~H\'ebrard\altaffilmark{12},
Coel~Hellier\altaffilmark{5},
Simon~T.~Hodgkin\altaffilmark{13},
Keith~D.~Horne\altaffilmark{2},
Stephen~R.~Kane\altaffilmark{14},
Pierre~F.~L.~Maxted\altaffilmark{5},
Andrew~J.~Norton\altaffilmark{10},
Don~L.~Pollacco\altaffilmark{15},
Frederic~Pont\altaffilmark{16},
Ian~Skillen\altaffilmark{17},
Barry~Smalley\altaffilmark{5},
Rachel~A.~Street\altaffilmark{18},
Stephane~Udry\altaffilmark{8},
Richard~G.~West\altaffilmark{19},
David~M.~Wilson\altaffilmark{5},
}

\affil{
\altaffilmark{1}Department of Physics, University of Warwick, Coventry CV4 7AL, UK\\ 
\altaffilmark{2}School of Physics and Astronomy, University of St Andrews, North Haugh, St Andrews, Fife KY16 9SS, UK\\
\altaffilmark{3}Department of Physics, University of Central Florida, Orlando, FL 32816-2385, USA\\
\altaffilmark{4}Department of Astronomy and Astrophysics, University of California, Santa Cruz, CA 95064, USA\\
\altaffilmark{5}Astrophysics Group, School of Chemistry and Physics, Keele University, Staffordshire, ST5 5BG, UK\\
\altaffilmark{6}Department of Physics, University of Oxford, Denys Wilkinson Building, Keble Road, Oxford OX1 3RH, UK\\
\altaffilmark{7}STScI, 3700 San Martin Drive, Baltimore, MD 21218, USA\\
\altaffilmark{8}Observatoire de Gen\`eve, Universit\'e de Gen\`eve, 51 Ch. des Maillettes, 1290 Sauverny, Switzerland\\
\altaffilmark{9}Institut d'Astrophysique et de Géophysique, Université de Liège, Allée du 6 Août, 17, Bat. B5C, Liège 1, Belgium\\
\altaffilmark{10}Department of Physics and Astronomy, The Open University, Milton Keynes, MK7 6AA, UK\\
\altaffilmark{11}Department of Physics and Astronomy, Vanderbilt
 University, Nashville, TN 37235, USA\\
\altaffilmark{12}Institut d'Astrophysique de Paris, UMR7095 CNRS, 
Universit\'e Pierre \& Marie Curie, 98bis boulevard Arago, 75014 Paris, 
France\\
\altaffilmark{13}Institute of Astronomy, Madingley Road, Cambridge CB3 0HA, UK\\
\altaffilmark{14}Michelson Science Center, Caltech, MS 100-22, 770 South Wilson Avenue, Pasadena, CA  91125, USA\\
\altaffilmark{15}Astrophysics Research Centre, School of Mathematics \&\ Physics, Queen's University, University Road, Belfast, BT7 1NN, UK\\
\altaffilmark{16}School of Physics, University of Exeter, Exeter, EX4 4QL, UK\\
\altaffilmark{17}Isaac Newton Group of Telescopes, Apartado de Correos 321, E-38700 Santa Cruz de la Palma, Tenerife, Spain\\
\altaffilmark{18}Las Cumbres Observatory, 6740 Cortona Dr. Suite 102,
Santa Barbara, CA 93117, USA\\
\altaffilmark{19}Department of Physics and Astronomy, University of Leicester, Leicester, LE1 7RH, UK\\
}
\email{P.J.Wheatley@warwick.ac.uk}
%
%
%




\begin{abstract}
We present a comparative study of the thermal emission of the transiting 
exoplanets WASP-1b and WASP-2b using the {\it Spitzer Space Telescope}. 
The two planets have very similar masses but suffer different levels of 
irradiation and are predicted to fall either side of a sharp transition between 
planets with and without hot stratospheres. WASP-1b is one of the most highly 
irradiated planets studied to date. We measure planet/star contrast ratios in 
all four of the IRAC bands for both planets (3.6--8.0\,\um), and our results
indicate the presence of a strong temperature inversion in the atmosphere of 
WASP-1b, particularly apparent at 8\um, and no inversion in WASP-2b. In both cases 
the measured eclipse depths favor models in which incident energy is not 
redistributed efficiently from the day side to the night side of the planet. 
We fit the \spitzer\  light curves simultaneously with the best available radial 
velocity curves and transit photometry in order to provide updated measurements 
of system parameters. We do not find significant eccentricity in the orbit of 
either planet, suggesting that the inflated radius of WASP-1b is unlikely to be 
the result of tidal heating. 
Finally, by plotting ratios of secondary eclipse depths at 8\um\ and 4.5\um\ 
against irradiation for all available planets, we find evidence for a sharp 
transition in the emission spectra of hot Jupiters at an irradiation level 
of $2\times 10^9\,\rm erg\,s^{-1}\,cm^{-2}$. We suggest this transition 
may be due to the presence of TiO in the upper atmospheres of the most 
strongly irradiated hot Jupiters.

\end{abstract}



\section{Introduction}


The {\it Spitzer Space Telescope} 
\citep{Werner04} has been used to carry out the 
first photometry and emission spectroscopy of 
exoplanets that orbit main sequence 
stars \citep{Deming05,Charbonneau05,Richardson07,Grillmair07}. 
This 
was
achieved by observing transiting planets at secondary 
eclipse (when the planet is eclipsed by the star) 
which allows the 
emission of the planet to be 
separated from that of the star. 
Photometry and spectroscopy are the key measurements needed to 
determine the physical properties of any astronomical object, and secondary 
eclipse observations allow us to consider the temperature structure and 
chemical composition of exoplanet atmospheres and the 
redistribution of energy from the day side to the night side of the planet 
\citep[e.g.][]{Burrows05,Fortney05,Seager05,Barman05}.

\begin{table*}
\begin{center}
\caption{Log of \spitzer\/  observations of WASP-1b and WASP-2b.}
\label{tab-log}
\begin{tabular}{crcccrrrrrc}
Target & Prog. & Date & Start time & Duration & Bary.\ corr. &\multicolumn{4}{c}{No.\ of frames $\times$ effective exposure per frame}&Pipeline\\
        &          &      & UTC        & s  & s      & 3.6\,$\rm \mu m$ & 4.5\,$\rm \mu m$ & 5.8\,$\rm \mu m$ & 8.0\,$\rm \mu m$ & version\\
& & & & & \\
WASP-1b & 30129 & 2007-09-08 & 14:02:56 & 27\,586 & $+238.2$ & $2072 \times 10.4$\,s & & $2072 \times 10.4$\,s & & S16.1.0\\
        &   282 & 2006-12-30 & 14:15:27 & 27\,146 & $+252.5$ & & $2055 \times 10.4$\,s & &$2055 \times 10.4$\,s & S15.0.5\\

WASP-2b & 30129 & 2007-07-01 & 13:44:18 & 12\,427 & $+240.3$ & $1818 \times \phantom{0}1.2$\,s & & $909 \times 10.4$\,s & & S16.1.0 \\
        &   282 & 2006-11-28 & 08:23:11 & 12\,032 & $-11.8$ & & $910 \times 10.4$\,s & & $910 \times 10.4$\,s & S15.0.5 \\
\end{tabular}
\end{center}
\end{table*}


Secondary eclipse detections have been made from the ground 
\citep[e.g.][]{Mooij09,Sing09} but since 
the
signal is weak even in the best cases 
(of order 0.1 per cent in hot Jupiters) the bulk of measurements to date 
are from space. 
Together with detections in the optical with \corot\ and \kepler\
\citep{Snellen09,Alonso09,Borucki09} 
and the near infrared with \hst\ \citep{Swain09a,Swain09b}, \spitzer\ is providing 
an increasingly clear
picture of the thermal emission of exoplanets. 


\spitzer\ detections of thermal emission have been reported in various combinations 
of wavebands between 3.6 and 24\,\um. The bulk of the newly-discovered hot 
Jupiters are detectable only in the shorter wavelength \spitzer\ bands of the IRAC 
instrument \citep[3.6, 4.5, 5.8 \& 8.0\,$\rm \mu m$;][]{Fazio04} but 
fortunately this is a range in which strong molecular bands are 
expected, providing good constraints on atmospheric conditions. The IRAC observing 
modes also allow more efficient observations of these fainter systems, to some 
extent compensating for their lower brightness. The 
rapidly growing number of moderately-bright transiting hot Jupiters 
therefore provide an excellent opportunity to improve our understanding of 
exoplanet atmospheres.


The existing \spitzer\ observations show that hot Jupiters are 
strongly heated by their parent stars, with typical brightness temperatures 
in the 
range 1000--2000\,K, and that their spectra deviate strongly from black bodies. 
Preliminary theoretical calculations predicted strong molecular absorption in 
the IRAC bands \citep{Burrows05,Fortney05,Seager05,Barman05}, 
for which there is evidence in some systems
\citep[e.g. HD189733b;][]{Charbonneau08}, 
however other systems 
have measured brightness temperatures in excess of expectations, indicating
emission features in the IRAC bands
\citep[e.g. HD209458b;][]{Knutson08}. 
In these cases it is thought that an 
opacity source high in the atmosphere results in a hot stratosphere and 
that the molecular bands are driven into emission by the 
temperature inversion 
\citep{Hubeny03, Fortney06, Harrington07, Burrows07, Sing08}. 
\citet{Fortney08} and \citet{Burrows08} suggest that 
irradiated exoplanets may fall into two distinct classes, those
with and those without hot stratospheres, depending on the level of incident 
stellar flux 
\citep[dubbed pM and pL class planets respectively by][]{Fortney08}. 
The brightest and best studied systems, 
HD209458b and HD189733b, fall either side of the predicted transition between 
these classes \citep{Fortney08}, and their IRAC fluxes support the 
presence of a hot stratosphere in 
HD209458b \citep{Burrows07,Knutson08} and its absence in 
HD189733b \citep{Charbonneau08}.
Most of the other systems that have been observed in all four IRAC bands also 
seem to support this overall picture \citep{Knutson09b,Machalek09,Todorov10,O'Donovan10,Machalek10,Campo10}, 
but XO-1b and TrES-3 do not, with XO-1b presenting evidence for a temperature 
inversion despite low irradiation \citep{Machalek08,Machalek09}, and TrES-3 
not exhibiting 
evidence for a temperature inversion despite high irradiation 
\citep{Fressin10}. 
It may be that additional parameters dictate the presence of a temperature 
inversion, although to some extent the picture is confused by the different models 
and criteria for inversion detection applied by different authors. \citet{Gillon10} 
show that the planets studied to date cover a wide region in color-color space, 
and do not fall clearly into two groups. 


In this paper we present \spitzer\ IRAC secondary eclipse detections of the 
transiting planets WASP-1b and WASP-2b, which were discovered by the 
Wide Angle Search for Planets (WASP) project \citep{Cameron07a,Pollacco06}.
These planets make an interesting pair for comparative study since 
they have near identical mass 
($0.9\,M_{\rm J}$) 
and yet WASP-1b is
highly irradiated and expected to have a hot stratosphere, 
while WASP-2b is not \citep{Fortney08}. Indeed, WASP-1b is one of the most highly 
irradiated planets studied with \spitzer\  to date (incident flux of $2.5\times10^9\,\rm erg\,s^{-1}\,cm^{-2}$).   
It is also one of the 
group of oversized hot Jupiters that have radii larger than can be 
explained with canonical models \citep{Charbonneau07}.

In addition to presenting and discussing \spitzer\ detections of WASP-1b and 
WASP-2b, we present revised parameters for both planets based on simultaneous 
fits of all available photometry and spectroscopy.

\begin{figure*}
\begin{center}
\plotone{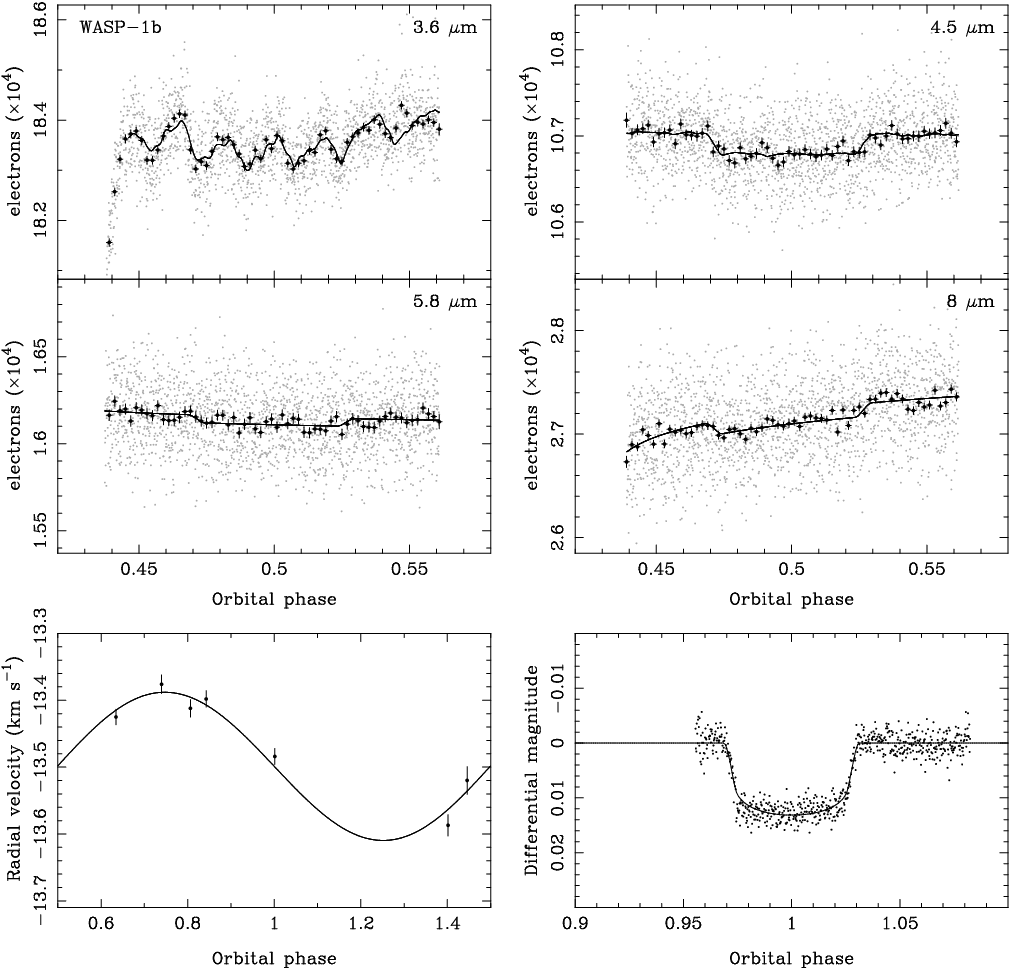}
\caption[]{
Upper panels show the \spitzer\ IRAC light curves of the star 
WASP-1 during the expected times of secondary eclipse of the planet WASP-1b. 
Points show the measurements from individual images, crosses show the data 
binned into five hundred bins per orbital period. Solid lines show our 
best fits to the eclipse light curves, including linear decorrelation and other 
instrumental effects described in Sects.~\ref{sec-insb} \& \ref{sec-sias}. 
The lower panels show the radial velocity curve of WASP-1 (measured with 
SOPHIE at OHP) and 
the z-band optical light curve during primary transit (from Keplercam), 
each plotted on different phase ranges. 
These data were fitted simultaneously with the 
secondary eclipse observations. 
}
\label{fig-w1-raw}
\end{center}

\end{figure*}
\begin{figure*}
\begin{center}
\plotone{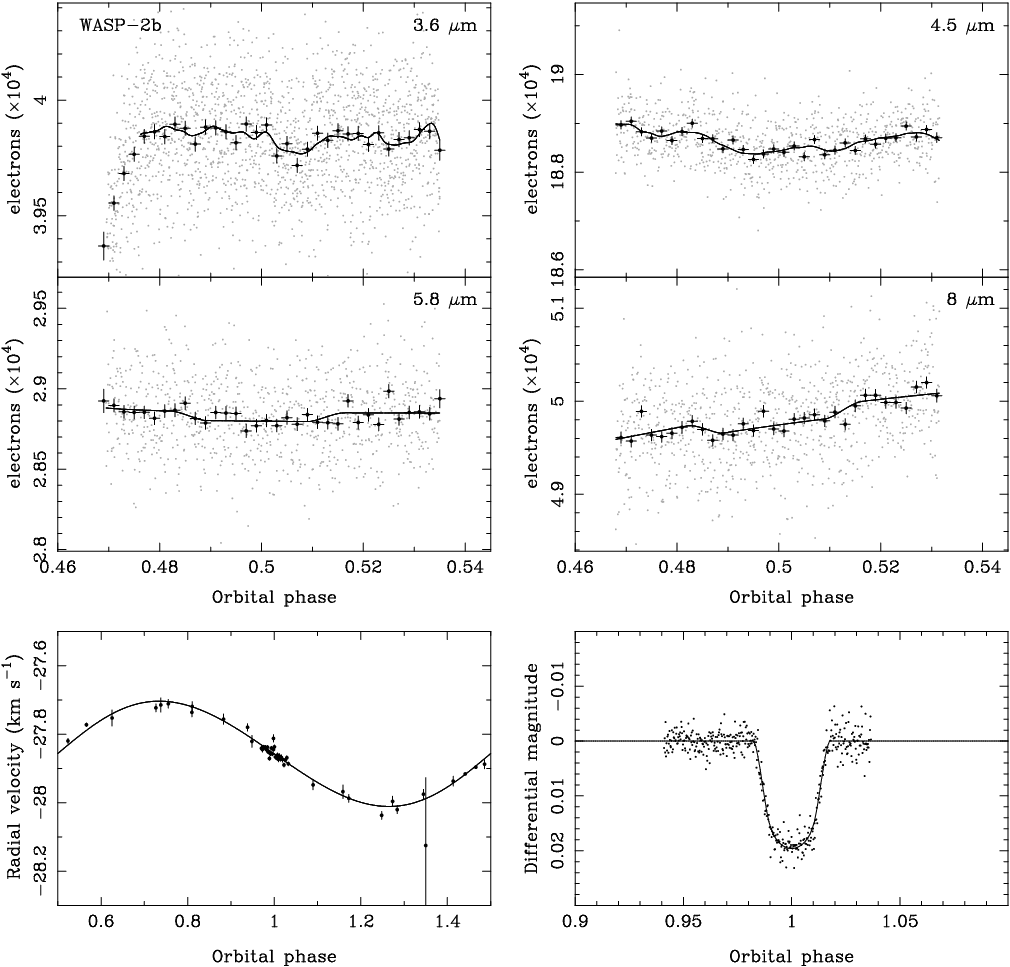}
\caption[]{
Upper panels show the \spitzer\ IRAC light curves of the star 
WASP-2 during the expected times of secondary eclipse of the planet WASP-2b. 
Points show the measurements from individual images, crosses show the data 
binned into five hundred bins per orbital period. Solid lines show our 
best fits to the eclipse light curves, including linear decorrelation and other 
instrumental effects described in Sects.~\ref{sec-insb} \& \ref{sec-sias}. 
The lower panels show the radial velocity curve of WASP-2 (measured with 
SOPHIE, Coralie and HARPS) and the z-band optical light curve during
primary transit (from Keplercam), each plotted on 
different phase ranges. These data were fitted simultaneously with the 
secondary eclipse observations. 
}
\label{fig-w2-raw}
\end{center}
\end{figure*}

\section{Observations}
\subsection{\spitzer\ secondary-eclipse observations}
The \spitzer\ IRAC instrument provides images of two adjacent fields, each in 
two wavebands
\citep[one field at 3.6 \& 5.8\,$\rm \mu m$, the other at 4.5 \& 8.0\,$\rm \mu m$;][]{Fazio04}. 
We observed the expected times of two secondary eclipses for each of 
WASP-1b and WASP-2b, allowing us to cover all four wavebands without 
repointing the telescope during an observation. 
A log of our four observations is given in Table\,\ref{tab-log}. 

The precise target positions were carefully chosen in order to avoid bad 
pixels, 
keep saturated stars off of the array, exclude bright stars from regions 
known to scatter light onto the IRAC 
detectors, and to place bright comparison stars on the detector (although 
these were not used in our final analysis). The pointing position was not 
dithered during the observations. 

Our observations were made in full array mode with 12\,s frame times (10.4\,s 
effective exposure), except for the 3.6/5.8\um\ observation of WASP-2b, which 
was made in stellar photometry mode with pairs of 2\,s frames taken in the 
3.6\um\ band for each 12\,s frame in the 5.8\um\ band. 
This mode was used in order 
to avoid saturating the target in the 3.6\um\ band. 
Observation durations are listed in Table\,\ref{tab-log} and were chosen 
in order to cover approximately twice the expected secondary eclipse 
duration \citep[transit durations are 3.7\,h and 1.8\,h for WASP-1b 
and WASP-2b respectively;][]{Charbonneau07}. 
%

\subsection{Radial velocity and transit observations}
In addition to the \spitzer\ secondary eclipse observations we also re-analyzed
radial-velocity measurements of WASP-1 and WASP-2 from \citet{Cameron07a} and 
$z$-band primary transit observations from \citet{Charbonneau07} as well as the 
SuperWASP discovery photometry from \citet{Cameron07a}. 

For WASP-1 we also 
included the $I$-band transit observations of \citet{Shporer07} in our analysis. 



For WASP-2 we included additional radial-velocity measurements made with the 
Swiss 1.2-m Euler telescope at La Silla, Chile between 2008 October 14 and 
November 6, and six pairs of measurements made with HARPS on the 
ESO 3.6-m telescope at La Silla on the nights of 2008 October 14/15, 15/16 
and 16/17.

All the radial-velocity measurements used in our analysis are plotted in the 
lower-left panels of Figs.\,\ref{fig-w1-raw}\,\&\,\ref{fig-w2-raw}. Primary transit 
photometry is shown in the lower-right panels of these figures, but we only
plot the transit photometry from \citet{Charbonneau07} for clarity. 



\section{\spitzer\ data reduction}
For our analysis we used the Basic Calibrated Data (BCD) files 
generated by the IRAC pipeline.\footnote{Pipeline description document: http://ssc.spitzer.caltech.edu/irac/dh/PDD.pdf} 
One image is produced for each exposure, 
with the processing including:
dark current and bias subtraction; 
{\em muxbleed} correction for the InSb arrays; 
non-linearity correction; scattered light subtraction; flat fielding; 
and photometric calibration. The pipeline version used to process each observation 
is given in Table\,\ref{tab-log}. 

We first renormalized each BCD image into units of electrons by multiplying 
by exposure time and detector gain and dividing by the flux conversion factor 
given in the BCD file headers. 

Light curves were extracted using simple aperture photometry with
a source aperture radius of four pixels in all bands. The aperture position was 
centred on the target in each image by centroiding the x and y pixel positions 
within a 9\,pix search box. 
The background contribution was estimated by 
calculating the mode of pixel values in an annulus centred on the 
source with inner and outer radii of 8 and 20 pixels. Uncertainties were 
estimated using counting statistics normalised to the measured 
variance in the background in each image. 

In order to remove contamination of the light curves by 
radiation hits
we adopted 
a two stage process in which we first rejected images for which photometric 
measurements were extreme outliers from the mean of the entire 
dataset ($>10\sigma$). We then formed a running mean of the remaining 
measurements across one hundred images and rejected images that were 
highly significant outliers from the running mean ($>5\sigma$). 
In total, 311 of 12,806 images were rejected (2 per cent). The proportion of 
frames rejected from each light curve is given in Table\,\ref{tab-pix}. 


The UTC times from the image headers were corrected to the Solar System 
barycenter using our own code and the co-ordinates of the \spitzer\ spacecraft 
from the 
HORIZONS on-line ephemeris system of the Jet 
Propulsion Laboratory.\footnote{http://ssd.jpl.nasa.gov/horizons.cgi}
The corrections applied to each observation are listed in Table\,\ref{tab-log}. 
The maximum change in correction during an observation was 2\,s, and so it was 
sufficient only to apply a single average offset to each observation. 

The resulting \spitzer\ light curves are plotted in 
Figs.\,\ref{fig-w1-raw}\,\&\,\ref{fig-w2-raw}. The light curves were fitted at 
full resolution (described in the following section), but they are 
plotted here also binned into five hundred phase bins for clarity.





\begin{figure*}
\begin{center}
\plottwo{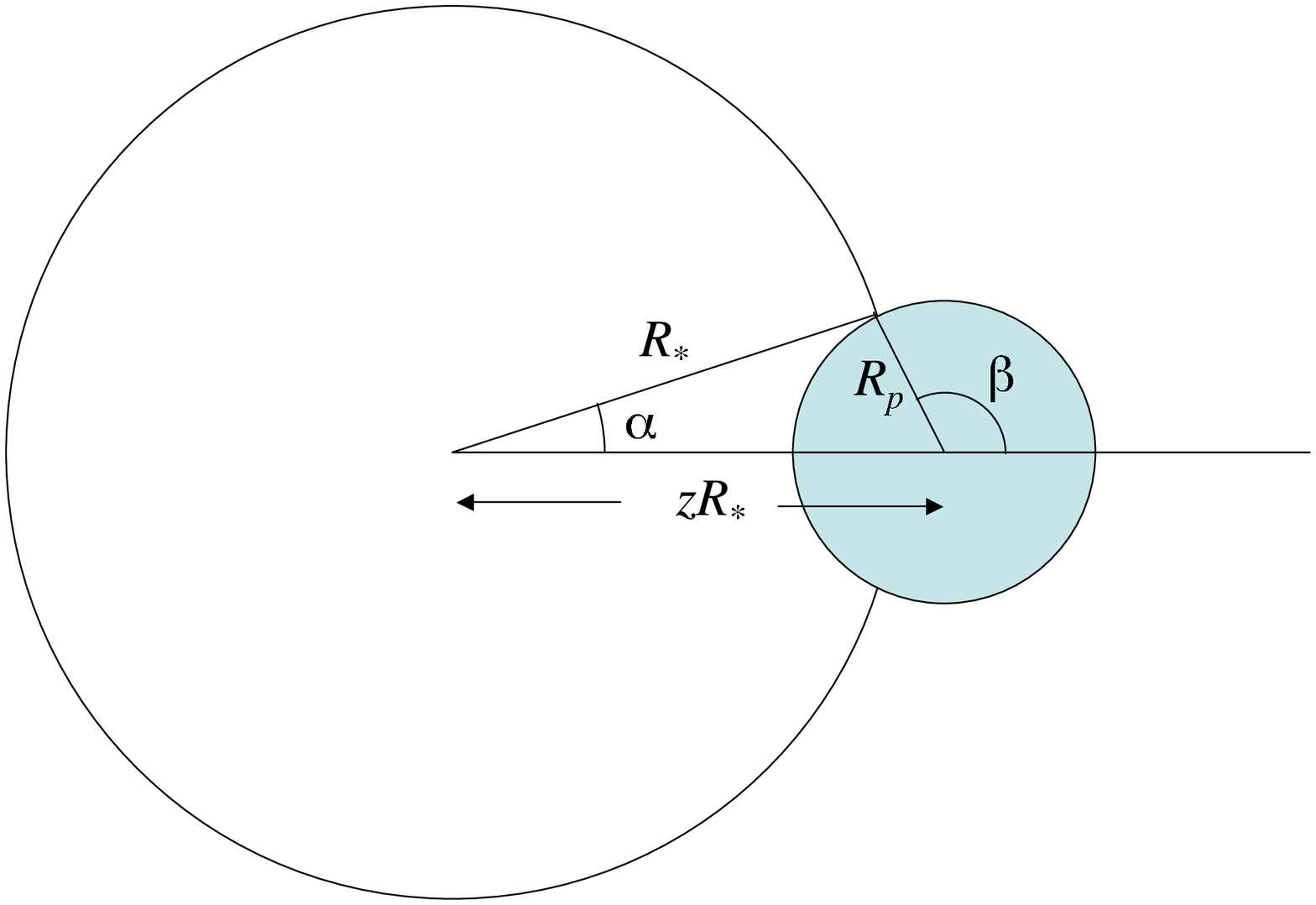}{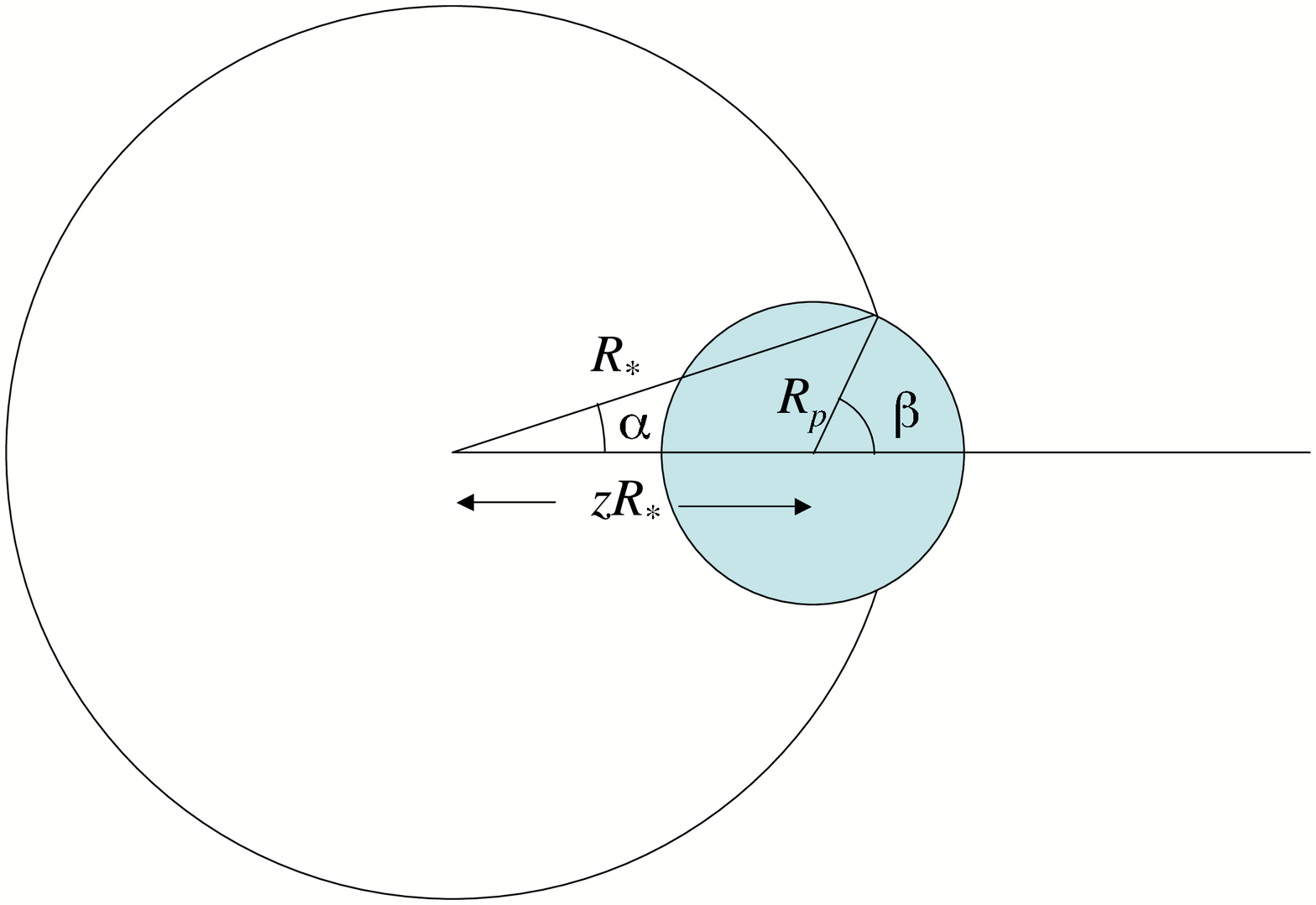}
\caption[]{Geometry of partial eclipse phases, showing the angles $\alpha$ and $\beta$ used in computing the visible fraction $\eta$ of the planetary disc. For illustrative clarity only, the planet is shown in front of the star.}
\label{fig:eclgeom}
\end{center}
\end{figure*}

\section{Light curve decorrelation and parameter fitting}

\label{sec-mcmc}

\subsection{3.6 and 4.5 $\rm \mu m$ decorrelation}

\label{sec-insb}

All four IRAC detectors exhibit distinctive patterns of correlated systematic 
error. Data from the 3.6 $\mu$m and 4.5 $\mu$m (InSb) detectors are most strongly 
affected by intra-pixel variations in quantum efficiency 
\citep{Reach05,Charbonneau05,Morales06}.
While the impact of these sensitivity differences is minimized by not moving the 
pointing during an observation,
the spacecraft pointing tends to oscillate around the nominal position, with an 
amplitude of around 0.1\,pix, leading to position-dependent variations in the 
measured stellar flux. The effect can be seen clearly in the 3.6\um\ light curve of 
WASP-1b (top left panel of Fig.\,\ref{fig-w1-raw}).
\citet{Knutson08} modelled these systematics using a two-dimensional polynomial 
fit to the instantaneous pointing offset from the centre of the pixel on 
which the stellar image was centred. For observations in which the pointing was 
changed during an observation it was found that a quadratic function was needed 
to remove this intra-pixel effect \citep{Knutson08,Charbonneau08}.
For observations where the pointing position is not changed
(like our own) some authors find acceptable fits 
using just a linear function of pixel position 
\citep{Knutson09b,Machalek09,Fressin10}.

\begin{table}
\begin{center}
\caption{Target pixel positions and fractions of rejected images for each of our \spitzer\ IRAC light curves. The pixel ranges correspond to the full range of the Gaussian-filtered positions used for decorrelation in Sect.\,\ref{sec-insb}.}
\label{tab-pix}
\begin{tabular}{ccrrrrc}
Target & Band & \multicolumn{2}{c}{x pixel} & \multicolumn{2}{c}{y pixel} & rejected \\
       &    \um     & median & range & median & range & \% \\
&&\\
WASP-1b & 3.6 & 192.14 & 0.15 & 152.44 & 0.22 & 2.1 \\
        & 4.5 & 187.14 & 0.15 &  94.72 & 0.21 & 1.8 \\
        & 5.8 & 185.51 &      & 153.57 &      & 3.5 \\ 
        & 8.0 & 186.26 &      &  94.09 &      & 3.7 \\
WASP-2b & 3.6 & 203.81 & 0.06 & 149.78 & 0.16 & 0.3 \\
        & 4.5 & 107.37 & 0.05 & 155.02 & 0.10 & 2.6 \\
        & 5.8 & 197.19 &      & 151.04 &      & 3.3  \\
        & 8.0 & 106.22 &      & 153.21 &      & 2.5 \\
\end{tabular}
\end{center}
\end{table}

For our analysis we tried both quadratic and linear decorrelation functions, 
similar to those of \citet{Knutson08} and \citet{Knutson09b}. The quadratic function 
is 
\begin{equation}
p'=1+c_x(x-x_0)+ c_{xx}(x-x_0)^2+c_y(y-y_0)+c_{yy}(y-y_0)^2+c_tt
\label{eqn-intrapix}
\end{equation}
where $(x, y)$ are the instantaneous coordinates of the stellar image, 
$(x_0, y_0)$ are the coordinates of the centre of the fiducial pixel, and $t$ is the
mid-time of the exposure. The last term allows for a linear trend in the 
light curves as noted by \citet{Knutson09b}.
The linear decorrelation function is the same as Eqn.\,\ref{eqn-intrapix},
except that the constants $c_{xx}$ and $c_{yy}$ were set to zero. 
The measured target positions were first smoothed using a moving 
Gaussian filter with a width of 12 observations. The median pixel positions and 
pixel ranges for each of our observations are given in Table\,\ref{tab-pix}. 

Despite the use of these decorrelation functions, we were unable to find 
acceptable fits when including the steep increase in flux seen at the beginning of 
the 3.6\um\ light curves of both planets 
(Figs.~\ref{fig-w1-raw} \& \ref{fig-w2-raw}). Inspection of the corresponding 
background light curves revealed variations similar 
to that noted by \citet{Knutson09b} and 
attributed to an illumination-dependent sensitivity effect similar to that seen 
in the 5.8 and 8.0\um\ detectors (see Sect.\,\ref{sec-sias}).
For the fits presented in this paper we decided to exclude the first 
30\,min and 21\,min respectively of the 3.6\um\ light curves of WASP-1b and WASP-2b.

\subsection{5.8 and 8.0 $\rm \mu m$ detrending}
\label{sec-sias}

The 5.6\um\ and 8.0\um\ (Si:As) detectors suffer from time-dependent sensitivity 
variations that depend on the recent illumination history of each pixel 
\citep{Knutson07}. This can be seen clearly in the 8\um\ light curves of both 
WASP-1b and WASP-2b (Figs.~\ref{fig-w1-raw} \& \ref{fig-w2-raw}). 
Again we followed the methodology of \citet{Knutson08} in fitting a quadratic ramp 
function in 
logarithmic time, of the form
\begin{equation}
p' = 1+c_t\tau+c_{tt}\tau^2,
\label{eqn-ramp}
\end{equation}
where $\tau=\ln(t-t_0)$, $t$ being the mid-time of the exposure and $t_0$ being a 
fiducial time 30 minutes prior to the first exposure of the observing sequence.

\begin{figure*}
\begin{center}
\plottwo{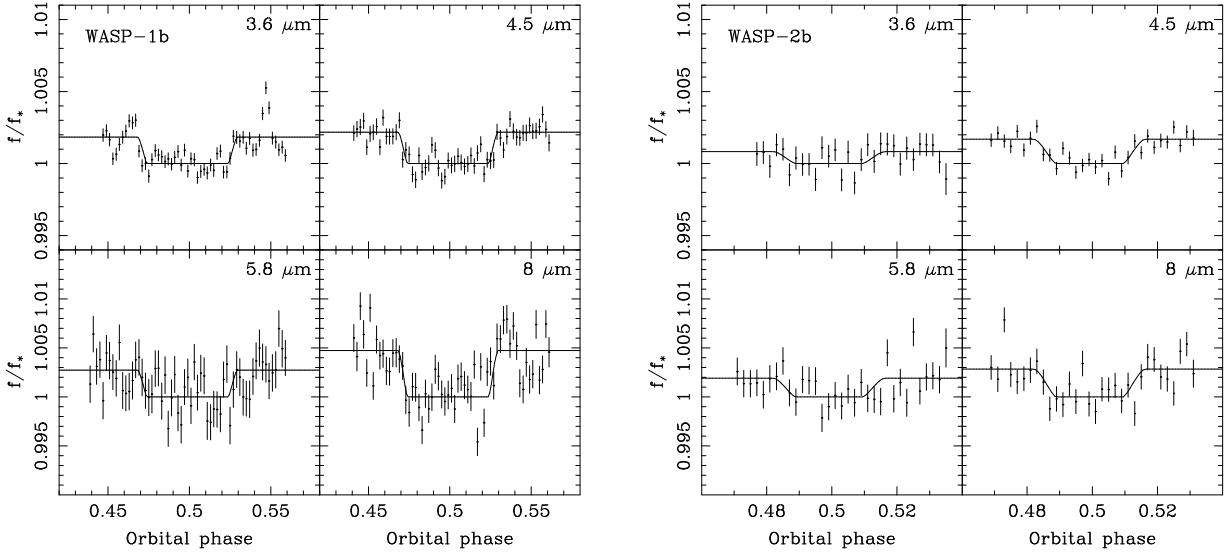}{wasp2_all_ecl_v4_lin.ps}
\caption[]{\spitzer\ IRAC light curves covering the secondary eclipses of the 
exoplanets WASP-1b and WASP-2b. Instrumental effects modelled in our fitting 
process have been removed and the light curves have been scaled to the flux of 
the star in each band. The light curves have been binned into five hundred 
phase bins per orbital period. 
}
\label{fig-ecl}
\end{center}
\end{figure*}

\begin{figure}
\begin{center}
\plotone{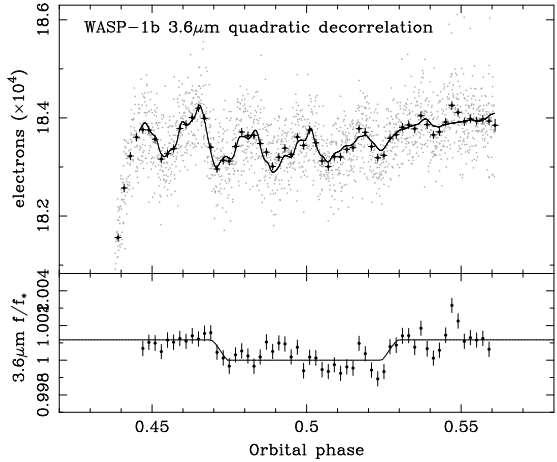}
\caption[]{The 3.6\um\ light curve of WASP-1b fitted using the quadratic 
decorrelation function described in Sect.~\ref{sec-insb}. The data have been 
binned into five hundred phase bins per orbital period. Instrumental effects have 
been removed in the bottom panel. 
}
\label{fig-quad}
\end{center}
\end{figure}

\subsection{Secondary eclipse profile}
\label{sec-lc}
The secondary eclipse profile is computed assuming the planet's day-side hemisphere to have a uniform surface brightness, and a star/planet flux ratio $f=F_p/F_*$. 

Given the ratio $p=R_p/R_*$ of the planetary to the stellar radius,
and a dimensionless separation $z$ in units of the stellar radius, the planet is partially eclipsed when $1-p<z<1+p$. At such times, the visible fraction of the planet's disk is given by
\begin{equation}
\eta(p,z)=\frac{\beta-\cos\beta\sin\beta+(\alpha-\cos\alpha\sin\alpha)/p^2}{\pi}
\label{eq:secvis}
\end{equation}
where
\begin{equation}
\cos\alpha=\frac{1-p^2+z^2}{2z}
\end{equation}
and
\begin{equation}
\cos\beta=\frac{1-p^2-z^2}{2pz},
\end{equation}
using the geometry sketched in Fig.~\ref{fig:eclgeom}. Outside transit, when $z>1+p$, the visible fraction of the planetary disk is $\eta = 1$. When $z<1-p$, the planet is totally eclipsed and $\eta=0$. At any time, the total observed flux from the star and planet is $F=F_*(1+f\eta(p,z))$.

\subsection{Parameter fitting}
\label{sec-param}

We solved for the full set of orbital and photometric parameters using the 
Markov-chain Monte-Carlo (MCMC) code described by \citet{Cameron07b} and 
\citet{Pollacco08}, modified to incorporate the secondary-eclipse and 
decorrelation models described in Sects.~\ref{sec-insb}--\ref{sec-lc}.

At each step in the Markov chain, synthetic optical light curves and 
radial-velocity curves are computed using the methodology described by 
\citet{Pollacco08}, using the first nine parameters listed in 
Table\,\ref{tab-param}. This particular set of parameters is chosen for 
their mutual near-orthogonality, as described by  \citet{Ford05} and 
\citet{Cameron07b}. The secondary eclipse models for the four IRAC 
detectors are computed as described in Sect.\,\ref{sec-lc}, 
multiplying the trial value 
of the planet/star flux ratios $f$ in each bandpass by the planet 
visibility function $\eta$ at each time of observation. The normalized 
light curve $(1+f\eta)$ is then co-multiplied by the appropriate 
sub-pixel (Eqn.\,\ref{eqn-intrapix}) or ramp (Eqn.\,\ref{eqn-ramp}) 
sensitivity model with a trial set of model 
coefficients 
$\{c_x,c_y,c_t\}$ (for linear decorrelation in the 3.6 \& 4.5\um\ bands),
$\{c_x,c_{xx},c_y,c_{yy},c_t\}$ (for quadratic decorrelation), or
$\{c_t,c_{tt}\}$ (in the 5.8 \& 8.0\um\ bands) for each detector in turn.

The normalized model light curve thus has the form 
\begin{equation}
p = (1+f\eta)(1+c_x(x-x_0)+c_y(y-y_0)+c_tt)
\end{equation}
for linear decorrelation of the 3.6\um\ and 4.5\um\ light curves,
\begin{equation}
p = (1+f\eta)(1+c_x(x-x_0)+ c_{xx}(x-x_0)^2+c_y(y-y_0)+c_{yy}(y-y_0)^2+c_tt)
\end{equation}
for quadratic decorrelation of the 3.6\um\ and 4.5\um\ detectors, and 
\begin{equation}
p = (1+f\eta)(1+c_t\tau+c_{tt}\tau^2)
\end{equation}
for the 5.8\um\ and 8.0\um\ light curves. 

The observed fluxes $d_i$ and the normalized model data $p_i$ are 
orthogonalized by subtracting their respective inverse variance-weighted 
mean values $\hat{d}$ and $\hat{p}$, then computing the inverse 
variance-weighted scale factor
\begin{equation}
F_*=\sum_i(d_i-\hat{d})(p_i-\hat{p})w_i/\sum_i(p_i-\hat{p})^2 w_i.
\end{equation}
where the weights $w_i=1/{\rm Var}(d_i)$ are the inverse variances associated 
with the observed fluxes. The logarithmic likelihood of obtaining the observed 
data given the model is quantified by 
\begin{equation}
\chi^2_{\rm Spitzer} = \sum_i w_i(d_i-\hat{d}-F_*(p_i-\hat{p}))^2.
\end{equation}
This contribution is added to the $\chi^2$ statistic computed for the photometric 
and radial-velocity data.  This is described in detail by \citet{Pollacco08}, 
to which the reader is referred.

\begin{table*}
\begin{center}
\caption{Proposal parameter values derived from MCMC parameter fitting to combined optical light curves, radial-velocity curves and Spitzer secondary-eclipse data.}
\label{tab-param}
\begin{tabular}{lccccr}
Parameter & Symbol & \multicolumn{2}{c}{WASP-1b} & WASP-2b & Units \\
          &        & linear decor. & quadratic decor. & linear decor. &  \\
&&\\

BJD of primary mid-transit & $T_0$ & 
$2453998.1924\pm0.0002$ &
$2453998.1924\pm0.0002$ &
$2454002.2754\pm0.0002$ &  d \\

Orbital period & $P$ &  
$2.519954\pm 0.000006$ & 
$2.519959\pm 0.000006$ & 
$2.152225\pm 0.000003$ & d \\

Eclipse duration & $t_T$ & 
$0.1550\pm 0.0006$ & 
$0.1550\pm 0.0006$ & 
$0.0752\pm 0.0009$ & d \\

Planet/star area ratio & $\Delta F$ & 
$0.0101\pm 0.0001$ & 
$0.0101\pm 0.0001$ & 
$0.0178\pm 0.0004$ & \\

Impact parameter & $b$ & 
$0.064^{+0.048}_{-0.042}$ & 
$0.044^{+0.052}_{-0.030}$ & 
$0.731^{+0.018}_{-0.022}$& \\

Stellar mass & $M_*$ & 
$1.217\pm 0.015$ & 
$1.216\pm 0.015$ & 
$0.861\pm 0.022$& M$_\odot$ \\

Stellar RV amplitude & $K$ & 
$0.111\pm 0.010$ & 
$0.111\pm 0.009$ & 
$0.154\pm 0.004$& km s$^{-1}$\\

& $e\cos\omega$ & 
$-0.0026\pm 0.0007$ & 
$-0.0012\pm 0.0007$ & 
$-0.0013\pm 0.0009$ & \\

& $e\sin\omega$ & 
$-0.0053\pm 0.0076$ & 
$-0.0083\pm 0.0073$ & 
$-0.048\pm 0.021$& \\

Flux ratio at 3.6 $\mu$m & $f_{3.6}$ & 
$0.00184\pm 0.00016$ & 
$0.00117\pm 0.00016$ & 
$0.00083\pm 0.00035$ & \\

Flux ratio at 4.5 $\mu$m & $f_{4.5}$ & 
$0.00217\pm 0.00017$ & 
$0.00212\pm 0.00021$ & 
$0.00169\pm 0.00017$ & \\

Flux ratio at 5.8 $\mu$m & $f_{5.8}$ & 
$0.00274\pm 0.00058$ & 
$0.00282\pm 0.00060$ & 
$0.00192\pm 0.00077$ & \\

Flux ratio at 8.0 $\mu$m & $f_{8.0}$ & 
$0.00474\pm 0.00046$ & 
$0.00470\pm 0.00046$ & 
$0.00285\pm 0.00059$ & \\

\end{tabular}
\end{center}
\end{table*}

At each step in the MCMC calculation, 
the planet/star flux ratios in the four IRAC bands 
and the nine other model parameters \{$T_0, P, t_T, \Delta F, b, K, e\cos\omega, e\sin\omega$\}
are each given
a random Gaussian perturbation, with a
"jump length" of order the uncertainty in the fitting parameter
concerned. The Spitzer secondary-eclipse data are then divided by the
scaled model secondary-eclipse profile, and the parameters of the
decorrelation function are fitted by linear least-squares, using
singular-value decomposition \citep{Press93}.
The procedure is the same as that used by \citet{Gillon10}.
The model data are computed and fitted to the observations, and the 
global $\chi^2$ statistic is computed. A decision is then made to either 
accept or reject the proposed parameter set according to the 
Metropolis-Hastings algorithm: if $\chi^2$ has decreased, the step 
is accepted unconditionally. If $\chi^2$ has increased by an 
amount $\Delta\chi^2$, the proposal may be accepted with 
probability $\exp(-\Delta\chi^2/2)$.

For the first few hundred steps of a typical run, the parameter values evolve 
in a way that drives $\chi^2$ down to a global minimum value. The algorithm 
then explores the joint posterior probability distribution of the parameters 
in the neighborhood of the optimal solution. We use the approach of 
\citet{Knutson08}, declaring the initial burn-in phase is complete when $\chi^2$ 
exceeds for the first time the median value of all previously-accepted values 
of $\chi^2$. At this point, we rescale the error estimates on the data points 
in each distinct photometric data set, such that the contribution of that data 
set to $\chi^2$ is approximately equal to the number of observations in the set. 
We then run the chain for a further few hundred steps, and compute new jump 
lengths for the individual fitting parameters from the chain variances. 
After a further short burn-in phase, the chain is allowed to continue for a 
production run of 30000 steps. The ensemble of models in this chain defines 
the joint posterior probability distribution for the full set of parameters.

The resulting set of fitting parameters, and the median and one-sigma errors 
of their posterior probability distributions, are listed in 
Table\,\ref{tab-param}, omitting only the rather lengthy list of 
decorrelation coefficients.


\section{Results}


The models best fitting our combined secondary-eclipse, primary-transit and 
radial-velocity data, using linear decorrelation in the 3.6\um\ and 4.5\um\ bands, 
are plotted in Figs.~\ref{fig-w1-raw} \& \ref{fig-w2-raw}.
The \spitzer\ data and 
best-fitting models are also plotted in Fig.~\ref{fig-ecl}, where the decorrelation 
functions have been removed for clarity. Fitted parameters are 
given in Table~\ref{tab-param}. In common with several recent studies, in which the
spacecraft pointing was not changed during an observation
\citep{Knutson09b,Machalek09,Fressin10}, 
we find that linear decorrelation provides generally a good description of the 
intra-pixel effect in the 3.6\um\ and 4.5\um\ bands. The exception is the 3.6\um\ 
light curve of WASP-1b, where some features possibly related to the intra-pixel 
effect remain. 

We fitted our data also using quadratic decorrelation functions 
in the 3.6\um\ and 4.5\um\ bands (Sect.~\ref{sec-insb}). 
This made no significant difference to the fits of any of our datasets apart from 
the 3.6\um\ band of WASP-1b, which is plotted 
in Fig.~\ref{fig-quad}. The fit in 
this band is improved by the quadratic decorrelation, and the eclipse depth is 
significantly reduced. The best-fit values for all parameters from this 
fit are given in Table~\ref{tab-param}.

\begin{figure*}
\begin{center}
\plottwo{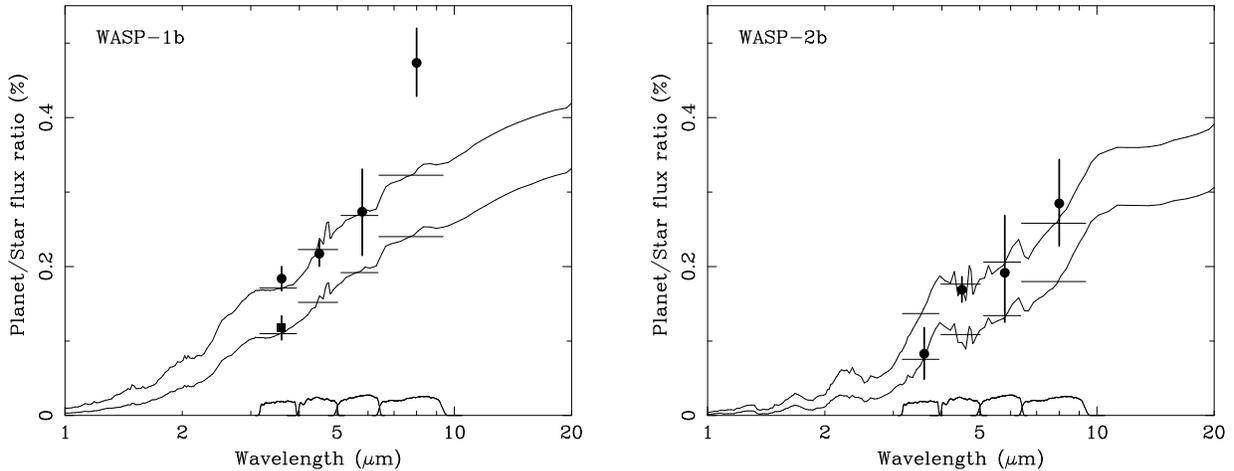}{wasp2_model_v4_lin.ps}
\caption[]{
Secondary eclipse depths of the exoplanets WASP-1b and WASP-2b 
overlaid on the planetary atmosphere models of \citet{Fortney08}, 
which are plotted as fraction of 
the stellar flux.
In both cases the upper curves represent models in which emission is only from the 
day-side of the planet, and the lower curves represent uniform emission 
from the 
entire exoplanet surface. 
For the 3.6\um\ band of WASP-1b we have included eclipse depths from both linear 
decorrelation (circle) and quadratic decorrelation (square). 
The horizontal 
lines on each model represent averages in the IRAC bands weighted by the 
spectral responses of the camera, which are indicated with solid curves 
at the bottom of the plot.  
}
\label{fig-model}
\end{center}
\end{figure*}

Our measured secondary-eclipse depths are compared with the planetary 
atmosphere models of \citet{Fortney08} in Fig.\,\ref{fig-model}.  
For WASP-1b the model predicts a stratospheric temperature inversion and the 
two curves represent calculations in which the planetary emission is 
either limited to the day side (upper curve) or is uniform over its entire 
surface (lower curve). 
For the 3.6\um\ band we include the eclipse depths from both the linear decorrelation
(circle) and the quadratic decorrelation (square). 
For WASP-2b, which is 
much less strongly irradiated, the model does not predict a temperature 
inversion and the two curves represent calculations for day-side emission 
(upper) and uniform emission (lower) as before. In order to aid comparison 
with measured eclipse depths we have calculated average planet/star flux ratios 
for each model in each IRAC band. These averages have been weighted using the 
IRAC spectral response curves \citep{Hora08}, which are plotted at the 
bottom of each panel. The model averages in each band are represented by 
horizontal lines. 

The 4.5\um\ and 5.8\um\ eclipse depths of WASP-1b are consistent with the model 
predicting a temperature inversion in this system and where emission is limited 
to the day side of the planet. The 5.8\um\ band is consistent also with uniform 
emission, but the 4.5\um\ band is not. The 3.6\um\ eclipse depth with linear 
decorrelation is also consistent with the inversion model and day-side emission, 
but the same data fitted with quadratic decorrelation is not. Since the 
quadratic decorrelation provides a better fit to the data 
(Fig.~\ref{fig-quad}), this suggests that the model is over-predicting the 
planetary flux in this band. Alternatively, it may be that the additional 
degrees of freedom in the quadratic decorrelation results in an underestimate 
of the eclipse depth in this band. 

The 8.0\um\ eclipse depth of 
WASP-1b lies significantly above the model prediction (3.3$\sigma$). 
The inversion model predicts emission from water in this band, so it may be 
that the model is under-predicting the strength of the temperature inversion in 
this system. 

In WASP-2b, the eclipse depths in all four bands are consistent with the model 
predicting no temperature inversion and with emission only from the day side 
of the planet. 
The 3.6\um\ and 5.8\um\ eclipses are consistent also with the uniform 
emission model, but the 8\um\ eclipse is only marginally consistent with this model 
and the 4.5\um\ eclipse is inconsistent (3.6$\sigma$).

%

\begin{figure}
\begin{center}
\plotone{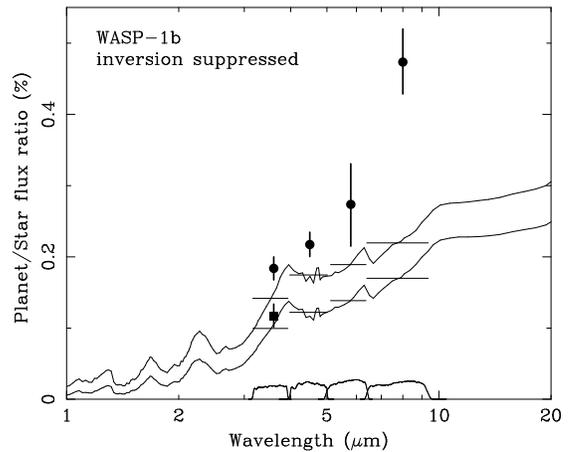}
\caption[]{
Secondary eclipse depths of the exoplanet WASP-1b overlaid on planetary atmosphere 
models in which the temperature inversion has been artificially suppressed by 
setting the TiO/VO abundance to zero. As in Figure\,\ref{fig-model}, the 
upper curve represents a model with emission only from the day side of the 
planet, and the lower curve represents a model in which heat is transported 
efficiently around the planet. 
For the 3.6\um\ band of WASP-1b we have included eclipse depths from both linear 
decorrelation (circle) and quadratic decorrelation (square). 
The horizontal lines show the weighted averages 
of the models in the four IRAC bands, with the IRAC band passes indicated by 
the solid curves at the bottom of the figure. 
}
\label{fig-noinv}
\end{center}
\end{figure}

As further illustration of the importance of the detected temperature inversion in 
WASP-1b, we have also compared the measured eclipse depths with a model calculation 
in which the temperature inversion has been artificially suppressed (by setting 
the TiO/VO abundances to zero in the model). The results are 
plotted in Fig.~\ref{fig-noinv} and show that the measured eclipse depths strongly 
favor the model with a temperature inversion.

Table~\ref{tab-param} includes updated system parameters for both planets 
from our simultaneous analysis of the secondary-eclipse, primary-transit and 
radial-velocity data. Of particular interest is $e\cos \omega$, which is 
constrained by the timing of the secondary eclipses. A significant eccentricity is 
found from the linear decorrelation fits to the WASP-1b light curves, however, 
this result is not supported by the quadratic decorrelation fits, where the 
deviation from zero drops to less than two-sigma. This fit places a tight 
three-sigma upper limit of $|e\cos\omega|<0.0033$, suggesting that the inflated 
radius of WASP-1b is unlikely to be due to tidal heating. We place a similar upper 
limit on the eccentricity of WASP-2b of $|e\cos\omega|<0.0040$. 

\section{Discussion}

Overall our measured secondary eclipse depths of WASP-1b and WASP-2b in the four 
IRAC bands are in remarkably good agreement with the predictions 
of the \citet{Fortney08} model (Fig.~\ref{fig-model}; note that the model has 
not been fitted to the data). This model predicts
an atmospheric temperature inversion in WASP-1b (class pM) and no inversion in 
WASP-2b (class pL). 
In WASP-2b the agreement between the model and data is excellent in all four bands. 
In WASP-1b the depth of the 8\um\ eclipse indicates 
that the strength of the temperature inversion may be under-estimated in the model. 
In both planets the 4.5\um\ eclipse is the best defined and also the one 
least affected by instrumental effects, and in both cases 
the eclipse depth in this band is consistent only with the day-side emission model. 
We take this as strong 
evidence that redistribution of incident energy is inefficient in both planets. 

Our detection of a temperature inversion in WASP-1b but not in WASP-2b is 
consistent with expectations based on their different irradiation levels \citep[$2.5\times10^9\,\rm erg\,s^{-1}\,cm^{-2}$ and $0.9\times10^9\,\rm erg\,s^{-1}\,cm^{-2}$ 
respectively;][]{Fortney08}.
It is also generally consistent with \spitzer\ measurements of other 
planets. Of the systems studied in all four IRAC bands, 
HD189733b, XO-2b and HAT-P-1b
have irradiation levels at or below the expected transition of around $0.8\times10^9\,\rm erg\,s^{-1}\,cm^{-1}$, 
and do not show strong temperature inversions, although XO-2b and HAT-P-1b do 
exhibit evidence for weak inversions
\citep{Charbonneau08,Machalek09,Todorov10}. 
HD209458b, TrES-4 and TrES-2 
have irradiation levels at or above the expected transition and all exhibit 
evidence for stronger temperature inversions 
\citep{Knutson08,Knutson09b,O'Donovan10}. 
However exceptions to this rule have also been found. 
XO-1b exhibits evidence for a temperature inversion despite 
irradiation at only $0.5\times10^9\,\rm erg\,s^{-1}\,cm^{-2}$ 
\citep{Machalek08,Machalek09}, while TrES-3 has not been found to do so, 
despite high irradiation at $1.6\times10^9\,\rm erg\,s^{-1}\,cm^{-2}$ 
\citep{Fressin10}. 

To some extent the observational picture is confused by the different models 
applied by different authors, and the different criteria applied for the detection 
of a temperature inversion. \citet{Gillon10} plotted a color-color diagram for 
planets studied in all four IRAC bands (their Fig.~10) and found that the 
planets studied to date cover a wide range of color space, and do not fall 
neatly into two distinct groups. In this paper we have presented a uniform 
analysis of secondary eclipse data from two planets for the first time. 
It may be that the observational picture will become more clear once further 
uniform analyses of multiple systems are carried out. 

\begin{figure} 
\begin{center}
\plotone{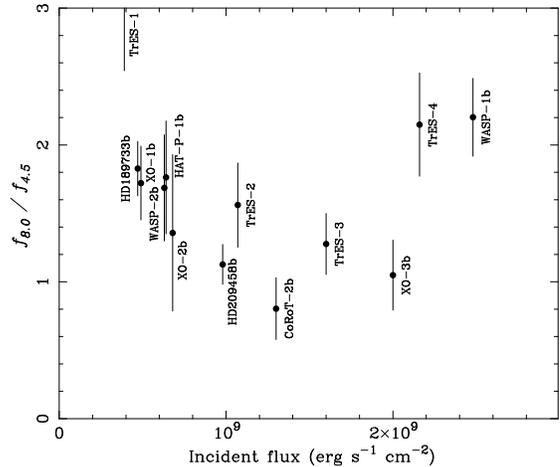}
\caption[]{The ratio of secondary-eclipse depths at 8\um\ and 4.5\um\ 
for all planets published to date (in planet/star contrast units), 
plotted as a function of irradiation by the 
parent star. 
The most highly irradiated planets, including WASP-1b, appear to be 
relatively brighter at 8\um. 
}
\label{fig-comp}
\end{center}
\end{figure}

In order to search for a pattern in published secondary eclipse depths, 
we have plotted the ratio of depths in the 8.0\um\ and 4.5\um\ bands against 
irradiation in Fig.~\ref{fig-comp}. We chose to compare the ratio of these two 
bands since they (1) are available for the largest number of planets, (2) are 
sensitive to the strength of water emission/absorption at 8\um, (3) are less 
affected by instrumental systematics than the 3.6\um\ band. 
Our figure does not show a sharp 
transition at an irradiation level of around 
$0.8\times10^9\,\rm erg\,s^{-1}\,cm^{-2}$ 
as might be expected from the 
predictions of \citet{Fortney08},
but it does show a gradual decrease of the flux ratio through this range. 
However, we do find tentative evidence for a sharp transition at a higher 
irradiation level of $2\times10^9\,\rm erg\,s^{-1}\,cm^{-2}$. 
The systems beyond this transition are TrES-4 \citep{Knutson09b} and WASP-1b 
(this paper), both showing remarkably strong emission at 8\um, perhaps 
related to their high irradiation levels. 

This transition
 could be due to a change in the chemistry and opacity of the atmospheres 
at that level of incident flux.  One possibility is that TiO gas is 
aloft at millibar pressures only at higher irradiation levels (and 
hotter atmospheres) and at lower irradiation 
($<2 \times 10^9\,\rm erg\,s^{-1}\,cm^{-2}$) 
another absorber is causing the observed inversions, such as a 
sulfur photochemical product 
\citep{Zahnle09}.
Both may operate over some range.  It now appears likely that 
opacity due to gaseous TiO and VO is not the entire story.  
In particular, the atmosphere of XO-1b appears to be so 
cold that Ti should be sequestered into solid condensates at 
any location in the atmosphere.  However, even at slightly 
warmer temperature in the upper atmosphere, there is the still 
the important issue of the temperature of the deep atmosphere.

Much attention has been paid to how cold traps may affect the 
abundance of TiO gas at millibar pressures.  If Ti is trapped in 
solid condensates deep in the atmospheres at $\sim$~0.1 to 1 kbar, 
then it would require quite strong vertical mixing to enable 
gaseous TiO to exist in the upper atmosphere.  This has been 
discussed by many authors 
\citep{Hubeny03,Fortney06,Fortney08,Burrows08}
and calculated in some detail by 
\citet{Showman09}
and especially 
\citet{Spiegel09}.
For instance, there is good agreement that if 
TiO causes the temperature inversion for HD209458b, it must be mixed 
above the cold trap 
\citep{Fortney08,Spiegel09}.
This is not a priori impossible.  In Neptune, abundant gaseous methane 
is found in the stratosphere, despite a cold trap that should condense 
methane in the troposphere 
\citep[][and references therein]{Fletcher10}.
However, it is not yet clear whether this mixing 
through the cold trap happens in hot Jupiters.

Interestingly, 
\citet{Fortney06}
investigated the deep atmosphere of HD149026b, which is at an irradiation 
level of $2 \times 10^9\,\rm erg\,s^{-1}\,cm^{-2}$ and found that its 
atmospheric temperatures sat on the cold trap dividing 
line (see their Fig.~2), which is a 
tantalizing suggestion that the break seen in Fig.~\ref{fig-comp} 
could be due to TiO.  
At higher irradiation levels there is no cold trap issue, 
although the exact location of this boundary is sensitive to the 
planet surface gravity and location of the radiative/convective boundary.  
We note that Spiegel et al., largely because they assumed colder interior 
temperatures, found that this division happened at higher irradiation 
levels.  Additional data for the hottest objects will help to support 
or refute the trends shown in Fig.~\ref{fig-comp}.

\section{Conclusions}
The measured secondary-eclipse depths of WASP-1b and WASP-2b presented here indicate
a strong temperature inversion in the atmosphere of WASP-1b, but no temperature 
inversion in WASP-2b. This difference is likely to be related to the 
much higher level 
of irradiation of the atmosphere of WASP-1b. The eclipse depths of 
both planets also favor models in which incident energy is not 
redistributed efficiently 
from the day side to the night side of the planet. We do not find significant 
eccentricity in the orbits of either planet, suggesting that the inflated radius of
WASP-1b is unlikely to have arisen through tidal heating. 
Finally, we find evidence for a sharp transition in the properties of 
planetary atmospheres with irradiation levels above 
$2\times10^9\,\rm erg\,s^{-1}\,cm^{-2}$, and we suggest that this transition 
might be due to the presence of TiO in the upper atmospheres of the most 
strongly irradiated hot Jupiters.



\acknowledgments



{\it Facilities:} \facility{Spitzer}.

\bibliographystyle{apj}
\bibliography{../../bibtex/refs2}

\clearpage

\end{document}